\documentclass{article}
\usepackage{spconf,amsmath,graphicx}
\usepackage{amsfonts} %

\usepackage{booktabs} %
\usepackage{tabularx}
\usepackage{multirow} %
\usepackage[T1]{fontenc}
\usepackage{setspace}
\usepackage{threeparttable}
\usepackage[labelfont=bf]{caption} %
\usepackage[bookmarks=true,hypertexnames=true,pagebackref=true]{hyperref}
\hypersetup{
  colorlinks=true, %
  urlcolor=black,   %
  linkcolor=blue,  %
  citecolor=red    %
}
\usepackage{color}
\usepackage[table]{xcolor}
\usepackage{cite}
\usepackage{circledsteps}
\usepackage{footnote}
\usepackage{arydshln}
\usepackage{soul}

\makeatletter
  \newcommand{\linkdest}[1]{\Hy@raisedlink{\hypertarget{#1}{}}}
\makeatother
\newcounter{mathsymbolcount}
\newcommand{\mathsymbol}[1]{%
  \ifcsdef{symbol:#1}{%
    {\hypersetup{hidelinks}\hyperlink{symbol:#1}{#1}}%
  }{%
    \stepcounter{mathsymbolcount}%
    \linkdest{symbol:#1}{}#1%
    \global\expandafter\def\csname symbol:#1\endcsname{}%
  }%
}

\newcolumntype{Y}{>{\centering\arraybackslash}X}

\renewenvironment{thebibliography}[1]{
  \begin{oldthebibliography}{#1}
  \setlength{\itemsep}{-0.1em}
  \setlength{\parskip}{0.em}
}
{
  \end{oldthebibliography}
}

\makeatletter
\def\bstctlcite{\@ifnextchar[{\@bstctlcite}{\@bstctlcite[@auxout]}}
\def\@bstctlcite[#1]#2{\@bsphack
  \@for\@citeb:=#2\do{%
    \edef\@citeb{\expandafter\@firstofone\@citeb}%
    \if@filesw\immediate\write\csname #1\endcsname{\string\citation{\@citeb}}\fi}%
  \@esphack}
\makeatother 

\newcommand{\newpara}[1]{\vspace{2pt}\noindent\textbf{#1}}

\title{Improving Design of Input Condition Invariant Speech Enhancement}
\name{Wangyou Zhang$^{1,2,}$\sthanks{$^*$Equal contribution.}, Jee-weon Jung$^{2,*}$, Shinji Watanabe$^2$, Yanmin Qian$^1$\thanks{The experiments were done using 1) the PI supercomputer at Shanghai Jiao Tong University supported in part by China STI 2030-Major Projects under Grant No. 2021ZD0201500, in part by China NSFC projects under Grants 62122050 and 62071288, and in part by Shanghai Municipal Science and Technology Commission Project under Grant 2021SHZDZX0102 and 2) the Bridges2 system at PSC and Delta systems at NCSA through allocation CIS210014 from the Advanced Cyberinfrastructure Coordination Ecosystem: Services \& Support (ACCESS) program, which is supported by National Science Foundation grants \#2138259, \#2138286, \#2138307, \#2137603, and \#2138296.}}
\address{$^1$Shanghai Jiao Tong University, China\quad$^2$Carnegie Mellon University, USA}

\begin{document}
\bstctlcite{IEEEexample:BSTcontrol} %
\ninept
\maketitle
\begin{abstract}
Building a single universal speech enhancement (SE) system that can handle arbitrary input is a demanded but underexplored research topic.
Towards this ultimate goal, one direction is to build a single model that handles diverse audio duration, sampling frequencies, and microphone variations in noisy and reverberant scenarios, which we define here as ``input condition invariant SE''.
Such a model was recently proposed showing promising performance; however, its multi-channel performance degraded severely in real conditions.
In this paper we propose novel architectures to improve the input condition invariant SE model so that performance in simulated conditions remains competitive while real condition degradation is much mitigated.
For this purpose, we redesign the key components that comprise such a system.
First, we identify that the channel-modeling module's generalization to unseen scenarios can be sub-optimal and redesign this module.
We further introduce a two-stage training strategy to enhance training efficiency.
Second, we propose two novel dual-path time-frequency blocks, demonstrating superior performance with fewer parameters and computational costs compared to the existing method.
All proposals combined, experiments on various public datasets validate the efficacy of the proposed model, with significantly improved performance on real conditions.
Recipe with full model details is released at \url{https://github.com/espnet/espnet}.
\end{abstract}

\begin{figure*}
  \centering
  \linkdest{figure:1}{}
  \includegraphics[width=0.88\textwidth]{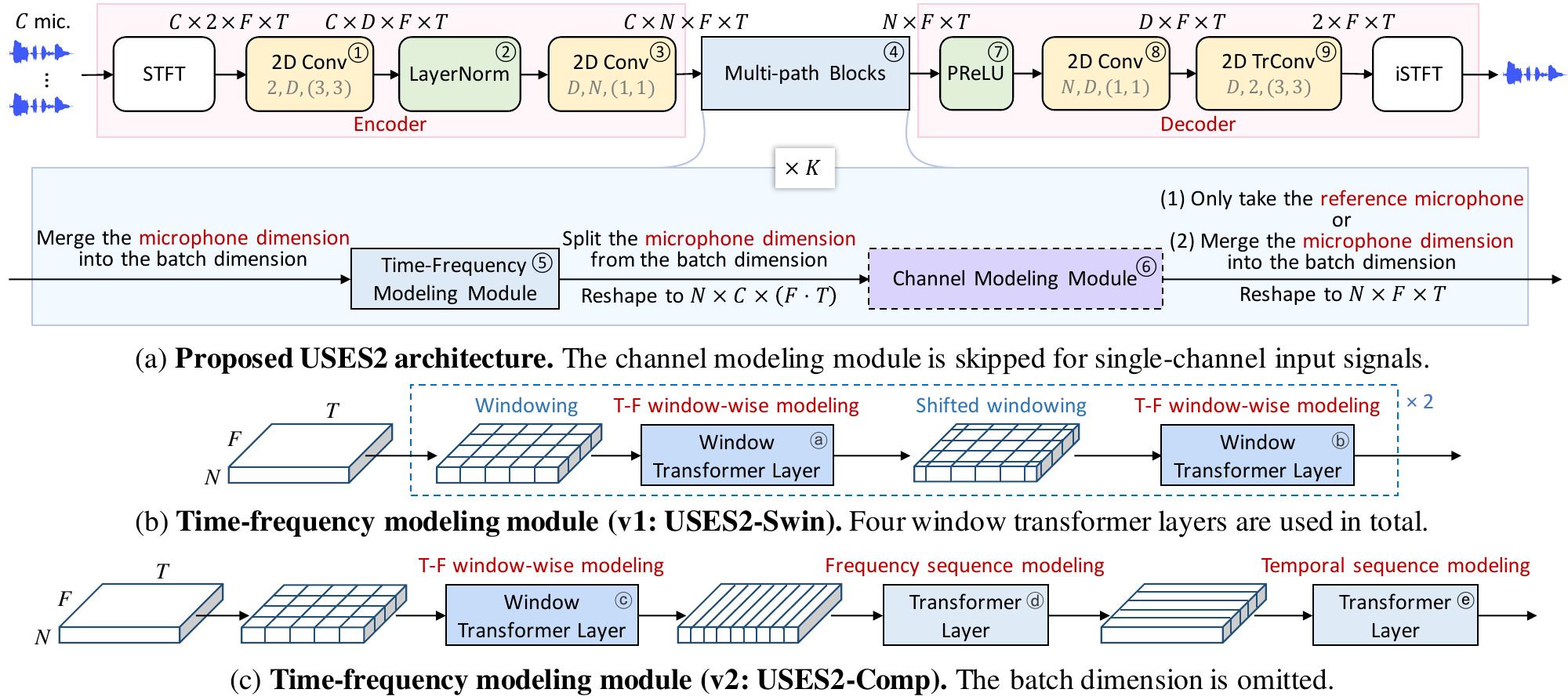}
  \captionsetup{labelfont=bf}
  \caption[overview]{The proposed {\protect\hypersetup{hidelinks}\hyperlink{symbol:input_condition_invariant}{input condition invariant}} SE model, USES2. Kernel size and feature maps of convolution layers are in gray.}
  \label{fig:overview}
  \vspace{-1.8em}
\end{figure*}
\vspace{-10pt}
\section{Introduction}
\label{sec:intro}
\vspace{-5pt}
Speech enhancement (SE) aims to remove undesired signals in noisy and reverberant environments, thus enhancing desired speech quality~\cite{Speech-Loizou2013}.
While SE can be divided into many subtasks, this paper primarily focuses on denoising and dereverberation as they are most common in real world scenarios.
The past decade has witnessed tremendous advancement of deep learning-based SE techniques with impressive performance in various scenarios. 
Spectral mapping~\cite{Regression-Xu2014}, time-frequency (T-F) masking~\cite{Training-Wang2014,Time_frequency-Williamson2017,Complex-Wang2020}, time domain approaches with learnable convolution-based models~\cite{Conv_TasNet-Luo2019,New-Pandey2019,DPRNN-Luo2020}, and sophisticated architectures~\cite{FRCRN-Zhao2022,TF_GridNet-Wang2023,Mask-Liu2023} all brought significant advances in the SE literature.
However, the majority of existing works mainly consider a single input condition, such as fixed single or multi-channel input, or input with a fixed sampling frequency.
Thus, the designed SE models cannot be directly used in different scenarios, e.g., when switching from one input condition to another.
The mismatch between input conditions is also known to degrade the model performance.

Recently, an unconstrained speech enhancement and separation network (USES)\footnote{``Unconstrained'' here means that the model is not constrained to be only used in a specific input condition.} has been proposed to handle diverse input conditions with a single model~\cite{Toward-Zhang2023}.
With careful design and integration of several key components, the USES model becomes independent of 1) the sampling frequency; 2) the number (and geometry) of input microphone channels; and 3) the input signal length.
The first independence is achieved leveraging the nature of short-time Fourier transform (STFT) and the adopted dual-path time-frequency model architecture~\cite{TFPSNet-Yang2022}.
The second is achieved by using the well-developed transform-average-concatenate (\mathsymbol{TAC}) technique~\cite{TAC-Luo2020}.
The last is achieved by introducing the memory tokens~\cite{Toward-Zhang2023}, which enable the model to perform segment-wise processing while maintaining the history information.
In this paper, we term this type of SE model as \linkdest{symbol:input_condition_invariant}{}``input condition invariant SE'', which is an important direction towards universal SE\,---\,our ultimate goal.

Although USES demonstrated promising performances on diverse datasets, we observe significant degradation in some real conditions, e.g., CHiME-4 multi-channel real recordings.
We analyze that this can be partly due to the sub-optimal architecture.
In particular, the adopted \mathsymbol{TAC} modules perform channel modeling by concatenating the channel-averaged representation to each channel-specific representation.
While this may be effective in many cases, it cannot well handle the microphone failure or channels with very different signal-to-noise ratios (SNRs).
In addition, the \mathsymbol{TAC} modules are used for processing both single- and multi-channel signals.
Such a coupled processing makes it difficult to tune the single- and multi-channel performance separately.
Hence, we argue that USES can be further developed to generalize better across various conditions.

In this paper, we aim to improve the model design of USES and enhance both performance and generalization in diverse conditions.
To this end, we propose to decouple the single-channel\,/\,multi-channel processing and redesign key components including new time-frequency modeling and channel modeling modules.
Based on the two proposed new architectures, {\hypersetup{hidelinks}\hyperlink{symbol:USES2-Comp}{USES2-Comp}} and {\hypersetup{hidelinks}\hyperlink{symbol:USES2-Swin}{USES2-Swin}}, we also adjust the technique combination accordingly to achieve the capability of handling diverse input conditions.
Additionally, we propose a two-stage training strategy to improve the training efficiency of the model, which allows improving the single- and multi-channel processing capabilities separately.
This also enables us to avoid data balancing between single- and multi-channel data.
Following~\cite{Toward-Zhang2023}, we experiment with the combination of five commonly-used public corpora (VoiceBank+DEMAND~\cite{Speech-Valentini-Botinhao2016}, DNS1~\cite{DNS_INTERSPEECH2020-Reddy2020}, CHiME-4~\cite{CHiME4-Vincent2017}, REVERB~\cite{REVERB-Kinoshita2013}, and WHAMR!~\cite{WHAMR-Maciejewski2019}) to demonstrate the model's capability.
Extensive results across datasets underscore the proposed single model's substantial improvement of SE performance and generalization, outperforming the conventional USES model in diverse conditions.

\vspace{-10pt}
\section{Proposed Methods}
\label{sec:proposed}
\vspace{-5pt}

\vspace{-10pt}
\subsection{Overview}
\label{ssec:overview}
\vspace{-5pt}
The proposed USES2 adopts the time-frequency dual-path modeling structure.
This makes the model \emph{inherently} sampling-frequency-independent (\mathsymbol{SFI}), identical to \cite{Toward-Zhang2023}.
As shown in Fig.~\ref{fig:overview} (a), the proposed model primarily comprises an STFT-based encoder {\hypersetup{hidelinks}\hyperlink{figure:1}{\Circled{\footnotesize 1}}}\,--\,{\hypersetup{hidelinks}\hyperlink{figure:1}{\Circled{\footnotesize 3}}}, a few multi-path blocks {\hypersetup{hidelinks}\hyperlink{figure:1}{\Circled{\footnotesize 4}}} for spectral and spatial modeling, and an iSTFT-based decoder {\hypersetup{hidelinks}\hyperlink{figure:1}{\Circled{\footnotesize 7}}}\,--\,{\hypersetup{hidelinks}\hyperlink{figure:1}{\Circled{\footnotesize 9}}}
Without loss of generality, the input signal has $\mathsymbol{C}$ microphone channels ($\mathsymbol{C} \geq$ 1).
The $\mathsymbol{C}$-channel signal is first converted to a complex-valued spectrum with shape $\mathsymbol{C}\times 2\times F\times T$, where $\mathsymbol{F}$ and $\mathsymbol{T}$ denote the frequency and temporal dimensions, respectively.
Each complex T-F bin in the spectrum is then projected to an $\mathsymbol{N}$-dimensional embedding by applying two 2D-convolution layers {\hypersetup{hidelinks}\hyperlink{figure:1}{\Circled{\footnotesize 1}}} {\hypersetup{hidelinks}\hyperlink{figure:1}{\Circled{\footnotesize 3}}} and a layer normalization {\hypersetup{hidelinks}\hyperlink{figure:1}{\Circled{\footnotesize 2}}}. 

The stacked $\mathsymbol{K}$ multi-path blocks {\hypersetup{hidelinks}\hyperlink{figure:1}{\Circled{\footnotesize4}}} perform the spectral and spatial modeling.
In the first $\mathsymbol{K_s}$ multi-path blocks, the feature is processed by both time-frequency modeling module {\hypersetup{hidelinks}\hyperlink{figure:1}{\Circled{\footnotesize5}}} and channel modeling module {\hypersetup{hidelinks}\hyperlink{figure:1}{\Circled{\footnotesize6}}}.
In the remaining $\mathsymbol{K}-\mathsymbol{K_s}$ blocks, for the sake of efficiency, only the time-frequency modeling module {\hypersetup{hidelinks}\hyperlink{figure:1}{\Circled{\footnotesize 5}}} is used for spectral refining.
The processed multi-channel feature is finally converted to single-channel by taking the reference channel and discarding the rest.
Note that we propose to always skip the channel modeling modules if the input is a single-channel signal, different to the design in~\cite{Toward-Zhang2023}.
This new \linkdest{symbol:decoupled_design}{}\textbf{decoupled design} allows separate optimization of single-channel and multi-channel processing capabilities, which cannot be achieved in~\cite{Toward-Zhang2023}.
The decoder with a parametric rectification linear unit (PReLU) {\hypersetup{hidelinks}\hyperlink{figure:1}{\Circled{\footnotesize7}}}, a point-wise convolution layer {\hypersetup{hidelinks}\hyperlink{figure:1}{\Circled{\footnotesize8}}}, and a a 2D transposed convolution layer (TrConv, {\hypersetup{hidelinks}\hyperlink{figure:1}{\Circled{\footnotesize9}}}) maps the feature to the complex-valued spectrum, which is further transformed to the waveform via iSTFT.

\vspace{-12pt}
\subsection{Time-frequency modeling module \Circled{\footnotesize 5}}
\label{ssec:tf}
\vspace{-5pt}
The original transformer-based frequency-path and time-path modeling in~\cite{TFPSNet-Yang2022,Toward-Zhang2023} exhibits strong performance. 
Nevertheless, they separately model the temporal and frequency dimensions sequentially, whereas joint modeling has the potential to be further beneficial~\cite{gu2020temporal,jung2022aasist}. %
Indeed, the short-time stationary nature of speech signals indicates that adjacent frames and frequency bins are highly correlated.
Therefore, exploiting the information in a local T-F region should also benefit the spectral modeling.
Inspired by the success of the Swin transformer in computer vision~\cite{Swin-Liu2021} and music classification~\cite{zhao2022s3t}, we incorporate the idea of transformer-based local window  modeling into SE.
Two novel architectures are proposed: (v1: {\hypersetup{hidelinks}\hyperlink{symbol:USES2-Swin}{USES2-Swin}}) purely Swin transformer-based joint T-F modeling; (v2: {\hypersetup{hidelinks}\hyperlink{symbol:USES2-Comp}{USES2-Comp}}) complementary T-F modeling.

\newpara{v1, Fig.~\ref{fig:overview} (b).} \mathsymbol{USES2-Swin} solely relies on the window-wise transformer layers for spectral modeling.
Similarly to the Swin transformer design~\cite{Swin-Liu2021,SwinIR-Liang2021}, the windowing and shifted windowing operations are respectively applied before the odd {\hypersetup{hidelinks}\hyperlink{figure:1}{\Circled{\footnotesize a}}} and even {\hypersetup{hidelinks}\hyperlink{figure:1}{\Circled{\footnotesize b}}} transformer layers.
In the windowing operation, the T-F representation is evenly partitioned into non-overlapping windows of size $W_F \times W_T$, where $\mathsymbol{W_F}$ and $\mathsymbol{W_T}$ are the numbers of frequency bins and frames in each window, respectively.
Zero-paddings are added before windowing (and later removed) so that the numbers of frequency bins and frames are exactly divisible by $\mathsymbol{W_F}$ and $\mathsymbol{W_T}$, respectively.
The window transformer layer is applied to each window independently for local T-F modeling.
Following~\cite{Swin-Liu2021}, a standard transformer structure~\cite{Attention-Vaswani2017} with layer normalization is used, except that a learnable 2D relative positional bias is added to the attention matrix in each multi-head self-attention module.
Thanks to the use of a fixed-duration STFT window, we have constant T-F resolution in the spectrum regardless of the sampling frequency (\mathsymbol{SF})~\cite{Toward-Zhang2023}\footnote{The T-F resolution here means the time\,/\,frequency difference between adjacent time frames\,/\,frequency bins.}.
Thus, the learned relative positional bias can be directly used without finetuning for handling signals of different \mathsymbol{SF}s.
Subsequently, the shifted windowing operation is introduced to expand the receptive field of each window in deeper layers.
It shifts the window partition by $\lfloor\frac{\mathsymbol{W_F}}{2}\rfloor$ and $\lfloor\frac{\mathsymbol{W_T}}{2}\rfloor$ along the frequency and time dimensions, respectively.
By stacking multiple transformer layers with windowing and shifted windowing, the T-F modeling module can effectively exploit information from different T-F regions.
(\linkdest{text:removal_mem}{}
Since this new structure can naturally handle arbitrarily long signals thanks to the window-based processing, we remove the memory tokens proposed in~\cite{Toward-Zhang2023} in this architecture.

\vspace{-2pt}
\newpara{v2, Fig.~\ref{fig:overview} (c).} \mathsymbol{USES2-Comp} regards the window-wise modeling as a complementary design to the existing frequency-path and time-path modeling.
We insert a window transformer layer {\hypersetup{hidelinks}\hyperlink{figure:1}{\Circled{\footnotesize c}}} at the beginning of each T-F modeling module to jointly capture the local T-F information.
Then, following~\cite{Toward-Zhang2023}, a group of learnable memory tokens \texttt{[mem]} of shape $1 \times \mathsymbol{N} \times 1 \times \mathsymbol{G}$ is concatenated before the processed feature along the time dimension for long sequence modeling, where $\mathsymbol{G}$ is the group size.
Two improved transformer layers~\cite{Dual_Path-Chen2020,TFPSNet-Yang2022} are subsequently used for frequency-path modeling {\hypersetup{hidelinks}\hyperlink{figure:1}{\Circled{\footnotesize d}}} and time-path modeling {\hypersetup{hidelinks}\hyperlink{figure:1}{\Circled{\footnotesize e}}}.

\vspace{-10pt}
\subsection{Channel modeling module \Circled{\footnotesize 6}}
\label{ssec:tac}
\vspace{-5pt}
In addition to the improved T-F modeling module, we further improve the channel modeling module {\hypersetup{hidelinks}\hyperlink{figure:1}{\Circled{\footnotesize 6}}} in Fig.~\ref{fig:overview} (a) to boost the spatial processing capability.
The vanilla \mathsymbol{TAC} module has limited capability in handling scenarios where different microphone channels have significantly different signal-to-noise ratios (SNRs) as observed in~\cite{Toward-Zhang2023}.
We conjecture that the design choice of the \mathsymbol{TAC} module incurs this limitation because it simply averages the representations from different channels.
The channel-wise attention~\cite{TPARN-Pandey2022} also handles different microphone numbers and permutations, achieving impressive performance in matched conditions.
However, this technique suffers from overfitting according to our preliminary experiments, violating our goal to build an {\hypersetup{hidelinks}\hyperlink{symbol:input_condition_invariant}{input condition invariant}} SE model with good generalization. (c.f. rows 5--6 in Table~\ref{tab:exp_chime4})

We propose a novel channel modeling module by combining the best of both approaches, named transform-attention-concatenate (\linkdest{symbol:TAttC}{}TA$_{\text{tt}}$C).
It maintains the channel independence and generalization while being capable of filtering noisy channels with attention.
Given an input feature $\mathbf{X} \in \mathbb{R}^{\mathsymbol{N} \times \mathsymbol{C} \times (\mathsymbol{F}\mathsymbol{T})}$, the channel modeling process of the proposed {\hypersetup{hidelinks}\hyperlink{symbol:TAttC}{TA$_{\text{tt}}$C}} module is formulated as follows:
{%
\setlength{\abovedisplayskip}{1pt plus 0pt minus 5pt}
\setlength{\belowdisplayskip}{1pt plus 0pt minus 5pt}
\begin{align}
    \mathbf{Y} &= \operatorname{PReLU}\big(\operatorname{FC}(\mathbf{X})\big) & \in \mathbb{R}^{H \times \mathsymbol{C} \times (\mathsymbol{F}\mathsymbol{T})} \,, \label{eq:transform} \\
    \bar{\mathbf{Y}} &= \operatorname{PReLU}\big(\operatorname{FC}\big(\operatorname{Attn}(\mathbf{Y})\big)\big) & \in \mathbb{R}^{H \times \mathsymbol{C} \times (\mathsymbol{F}\mathsymbol{T})}\,, \label{eq:attention} \\
    \hat{\mathbf{X}} &= \operatorname{LN}\Big(\operatorname{PReLU}\big(\operatorname{FC}([\mathbf{Y}, \bar{\mathbf{Y}}])\big)\Big) & \in \mathbb{R}^{\mathsymbol{N} \times \mathsymbol{C} \times (\mathsymbol{F}\mathsymbol{T})}\,, \label{eq:concatenate}
\end{align}
}where $[\mathbf{Y}, \bar{\mathbf{Y}}] \in \mathbb{R}^{2H \times \mathsymbol{C} \times (\mathsymbol{F}\mathsymbol{T})}$ denotes concatenating two features along the embedding dimension.
$\mathsymbol{H}$ is the projected embedding dimension.
$\operatorname{\mathsymbol{FC}}(\cdot)$ and $\operatorname{\mathsymbol{LN}}(\cdot)$ respectively denote the linear projection and layer normalization, both operating along the embedding dimension.
$\operatorname{Attn}(\cdot)$ denote the channel-wise attention as defined below:
{%
\setlength{\abovedisplayskip}{1pt plus 0pt minus 5pt}
\setlength{\belowdisplayskip}{1pt plus 0pt minus 5pt}
\begin{align}
    \mathbf{Q} &= \operatorname{Reshape}\!\Big(\!\operatorname{\mathsymbol{LN}}\!\Big(\!\operatorname{ReLU}\big(\operatorname{FC}(\mathbf{Y})\big)\Big)\Big) \in \mathbb{R}^{1 \times \mathsymbol{C} \times (\mathsymbol{H}\mathsymbol{F}\mathsymbol{T})} \,, \label{eq:attention_q} \\
    \mathbf{K} &= \operatorname{Reshape}\!\Big(\!\operatorname{\mathsymbol{LN}}\!\Big(\!\operatorname{ReLU}\big(\operatorname{\mathsymbol{FC}}(\mathbf{Y})\big)\Big)\Big) \in \mathbb{R}^{1 \times \mathsymbol{C} \times (\mathsymbol{H}\mathsymbol{F}\mathsymbol{T})} \,, \label{eq:attention_k} \\
    \mathbf{V} &= \operatorname{\mathsymbol{LN}}\Big(\operatorname{ReLU}\big(\operatorname{\mathsymbol{FC}}(\mathbf{Y})\big)\Big) \in \mathbb{R}^{\mathsymbol{H} \times \mathsymbol{C} \times (\mathsymbol{F}\mathsymbol{T})} \,, \label{eq:attention_v} \\
    \mathbf{A} &= \operatorname{\mathsymbol{LN}}\left(\operatorname{ReLU}\left(\operatorname{FC}\left(\operatorname{Softmax}\left(\frac{\mathbf{Q}\mathbf{K}^\textsf{T}}{\sqrt{\mathsymbol{H}\mathsymbol{T}^2}}\right) \mathbf{V}\right)\right)\right) \,, \label{eq:attention_a}
\end{align}
}where $\mathbf{A} \in \mathbb{R}^{\mathsymbol{H} \times \mathsymbol{C} \times (\mathsymbol{F}\mathsymbol{T})}$ is the output of the $\operatorname{Attn}$ module.
Note that the $\operatorname{Attn}$ module merges the time and frequency dimensions into the embedding dimension when calculating the attention map $\operatorname{Softmax}\left(\frac{\mathbf{Q}\mathbf{K}^\textsf{T}}{\sqrt{\mathsymbol{H}\mathsymbol{T}^2}}\right) \in \mathbb{R}^{1\times \mathsymbol{C}\times \mathsymbol{C}}$ in Eq.~(\ref{eq:attention_a}), which makes it independent of the size of $\mathsymbol{F}$ and $\mathsymbol{T}$.

\vspace{-12pt}
\subsection{Two-stage training strategy}
\label{ssec:strategy}
\vspace{-5pt}
We propose a two-stage training strategy to improve the training efficiency of the proposed SE model.
Since we have {\hypersetup{hidelinks}\hyperlink{symbol:decoupled_design}{decoupled}} the single- and multi-channel processing in Section~\ref{ssec:overview}, it is now possible to separate the optimization of these two capabilities with single- and multi-channel data, respectively.
Specifically, in the first stage, we update all parameters except for the {\hypersetup{hidelinks}\hyperlink{symbol:TAttC}{TA$_{\text{tt}}$C}} modules {\hypersetup{hidelinks}\hyperlink{figure:1}{\Circled{\footnotesize 6}}} in Fig.~\ref{fig:overview} (a) on diverse single-channel data.
This allows the model to focus on improving the single-channel SE performance.
In the second stage, we train the {\hypersetup{hidelinks}\hyperlink{symbol:TAttC}{TA$_{\text{tt}}$C}} modules {\hypersetup{hidelinks}\hyperlink{figure:1}{\Circled{\footnotesize 6}}} while freezing all other parameters on diverse multi-channel data.
This ensures to improve the multi-channel SE performance on top of the well-trained single-channel SE function with better generalization.
The benefits of the proposed two-stage training strategy include:
1) it is more flexible as we can use single- and multi-channel data separately for training without caring about the data balance issue;
2) we can optimize multi-channel processing independently without affecting single-channel SE capabilities, which is preferable in some practical applications.

\begin{table}
    \setstretch{0.92}
    \captionsetup{labelfont=bf}
    \caption{Information of the datasets. 
    ``\#Ch'' denotes the number of microphone channels.
    ``(Sim)'' and ``(Real)'' denote the synthetic and recorded data. 
    ``(A)'' and ``(R)'' represent anechoic and reverberant.}
    \label{tab:corpora}
    \centering
    \resizebox{1.0\linewidth}{!}{%
        \begin{tabular}{l|ccc} 
        \toprule
        \textbf{Dataset} & \textbf{Hours (train\,/\,dev\,/\,test)}  & \textbf{\mathsymbol{SF}} & \textbf{\#Ch}  \\
        \midrule
        \href{https://datashare.ed.ac.uk/handle/10283/2791}{VoiceBank+DEMAND}~\cite{Speech-Valentini-Botinhao2016} & 8.8\,\textbf{/}\,0.6\,\textbf{/}\,0.6 & 48 kHz & 1\\
        \href{https://github.com/microsoft/DNS-Challenge/tree/interspeech2020/master}{DNS1} (v1)~\cite{DNS_INTERSPEECH2020-Reddy2020} & (A)90\,\textbf{/}\,(A)10\,\textbf{/}\,(R)0.42\,\&\,(A)0.42 & 16 kHz & 1 \\
        \href{http://spandh.dcs.shef.ac.uk/chime\_challenge/chime2016/}{CHiME-4}~\cite{CHiME4-Vincent2017} & (Sim)14.7\,\textbf{/}\,(Sim)2.9\,\textbf{/}\,(Sim)2.3\,\&\,(Real)2.2 & 16 kHz & 5\\
        \href{https://reverb2014.dereverberation.com}{REVERB}~\cite{REVERB-Kinoshita2013} & (Sim)15.5\,\textbf{/}\,(Sim)3.2\,\textbf{/}\,(Sim)4.8\,\&\,(Real)0.7 & 16 kHz & 8\\
        
        \cellcolor[HTML]{EEEEEE} & \cellcolor[HTML]{EEEEEE}(R)58.0\,\textbf{/}\,(R)14.7\,\textbf{/}\,(R)9.0 & \cellcolor[HTML]{EEEEEE} & \cellcolor[HTML]{EEEEEE} \\
        \cellcolor[HTML]{EEEEEE}\multirow{-2}{*}{\href{https://wham.whisper.ai}{WHAMR!}~\cite{WHAMR-Maciejewski2019}} & \cellcolor[HTML]{EEEEEE}(A)58.0\,\textbf{/}\,(A)14.7\,\textbf{/}\,(A)9.0 & \cellcolor[HTML]{EEEEEE}\multirow{-2}{*}{16 kHz} & \cellcolor[HTML]{EEEEEE}\multirow{-2}{*}{2} \\
       \bottomrule
       \end{tabular}%
    }
\end{table} %
\begin{table*}
    \setstretch{0.9}
    \caption{Experiments on the proposed methods trained and evaluated using the CHiME-4 (5ch) dataset. 
    Models are trained only on 8 kHz data, and tested on 16 kHz data. 
    The values within parentheses denote single-channel performance while all others are multi-channel performance.
    The best and second best results are made bold and \underline{underlined}. All models except for Nos. 1 and 2 are our novel explorations.}
    \label{tab:exp_chime4}
    \centering
    \setlength{\tabcolsep}{3pt}
    \resizebox{\textwidth}{!}{%
    \begin{tabular}{llccc|ccccc|cc}
        \toprule
        \multirow{2}{*}{\textbf{No.}} & \multirow{2}{*}{\textbf{Model}} & \multirow{2}{*}{\textbf{\#Param}} & \multicolumn{2}{c|}{\textbf{\#MAC (G/s)}} & \multicolumn{5}{c|}{\textbf{Test (CHiME-4 Simu)}} & \multicolumn{2}{c}{\textbf{Test (CHiME-4 Real)}} \\
        & & & \textbf{1ch} & \textbf{2ch} & \textbf{PESQ-WB} $\uparrow$ & \textbf{STOI ($\times100$)} $\uparrow$ & \textbf{SDR (dB)} $\uparrow$ & \textbf{DNSMOS OVRL} $\uparrow$ & \textbf{WER (\%)} $\downarrow$ & \textbf{DNSMOS OVRL} $\uparrow$ & \textbf{WER (\%)} $\downarrow$ \\
        \hline
        1 & No processing & - & - & - & 1.27 & 87.0 & 7.5 & 2.08 & 5.8 & 1.46 & \textbf{6.7} \\
        2 & USES ({\em baseline})~\cite{Toward-Zhang2023} & 3.05 M & 65.3 & 98.0 & \underline{3.16} & \underline{98.3} & 20.6 & \underline{3.22} & 4.2 & 1.99 (2.94) & 78.1 (11.0) \\
        \hline
        3 & w/ {\hypersetup{hidelinks}\hyperlink{symbol:decoupled_design}{decoupled}} proc. & 3.05 M & 60.8 & 98.0 & 2.62 & 97.0 & 18.2 & 3.08 & 5.2 & 2.32 & 27.8 \\
        4 & No.3 + 2-stage training & 3.05 M & 60.8 & 98.0 & 2.23 & 94.9 & 15.6 & 2.99 & 6.8 & 2.41 & 23.8 \\
        5 & No.3 + Att$\times1$ + \mathsymbol{TAC}$\times2$ & 3.02 M & 60.8 & 97.5 & \textbf{3.21} & \textbf{98.5} & \textbf{22.2} & 3.20 & \textbf{4.0} & 1.48 & 99.0 \\
        6 & No.5 + 2-stage training & 3.02 M & 60.8 & 97.5 & 2.98 & 97.9 & 19.0 & \underline{3.22} & 4.4 & 1.58 & 93.8 \\
        7 & No.3 +  {\hypersetup{hidelinks}\hyperlink{symbol:TAttC}{TA$_{\text{tt}}$C}}$\times3$ & 3.47 M & 60.8 & 116.9 & 2.72 & 97.3 & 18.7 & 3.14 & 5.2 & 2.40 & 28.8 \\
        8 & No.7 + 2-stage training & 3.47 M & 60.8 & 116.9 & 3.06 & 98.2 & 19.8 & 3.21 & 4.2 & 2.40 & 54.3 \\
        \hline
        9 & \mathsymbol{USES2-Swin} ({\em proposed}) & 2.92 M & 37.7 & 75.5 & 2.94 & 98.1 & 20.6 & 3.21 & 4.2 & 2.84 & 22.1 \\
        10 & No.9 + 2-stage training & 2.92 M  & 37.7 & 75.5 & 2.98 & 98.2 & \underline{21.1} & \underline{3.22} & \underline{4.1} & 2.80 & 24.9 \\
        11 & \mathsymbol{USES2-Comp} ({\em proposed}) & 2.53 M & 52.4 & 83.0 & 3.05 & \underline{98.3} & 20.4 & 3.21 & 4.2 & \underline{2.89} & 15.6 \\
        12 & No.11 + 2-stage training & 2.53 M & 52.4 & 83.0 & 2.94 & 97.9 & 18.8 & \textbf{3.23} & 4.6 & \textbf{2.96} & \underline{12.1} \\
        \bottomrule
    \end{tabular}%
    }
    \vspace{-8pt}
\end{table*} %
\begin{figure*}
  \vspace{2.5pt}
  \centering\includegraphics[width=\textwidth]{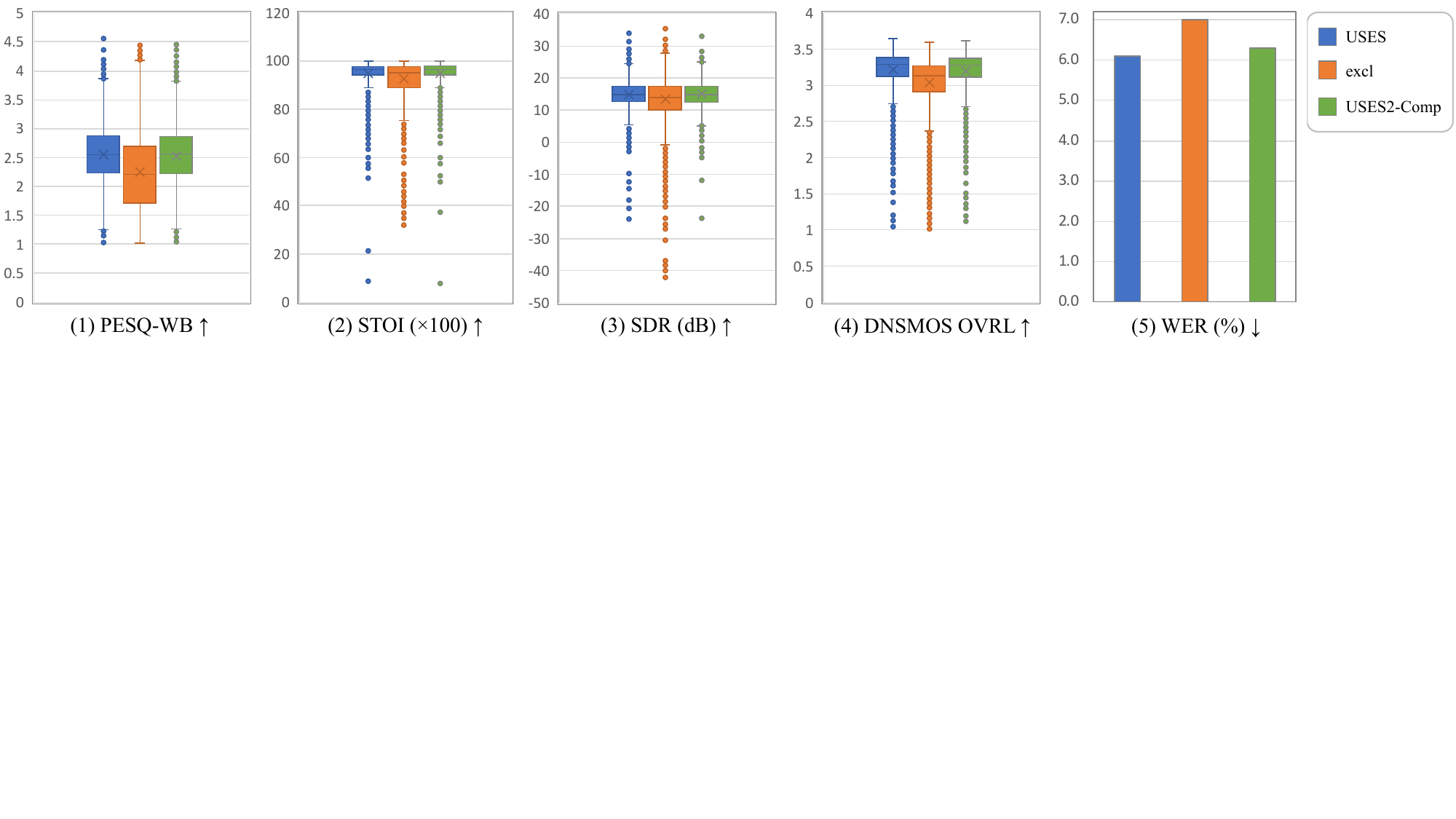}
  \captionsetup{labelfont=bf}
  \caption[simu_results]{Speech enhancement and recognition performance in diverse simulated conditions, averaged over the five corpora listed in Table~\ref{tab:corpora}. The models are trained only on 8 kHz data (via downsampling), and tested with the original sampling frequencies. ``USES'', ``\mathsymbol{excl}'', and ``U2-C'' denote the SE model proposed in~\cite{Toward-Zhang2023}, corpus-exclusive SE, and the proposed \protect\mathsymbol{USES2-Comp} model, respectively. We follow the same training and evaluation configurations as in~\cite{Toward-Zhang2023} to make the results comparable.}
  \label{fig:simu_results}
  \vspace{-1.8em}
\end{figure*}

\vspace{-10pt}
\section{Experiments}
\label{sec:exp}
\vspace{-10pt}

\subsection{Datasets}
\label{ssec:db}
We configure a global training set spanning {\raise.17ex\hbox{$\scriptstyle\sim$}}245 hours to train our USES2 combining five well recognized datasets, identical to \cite{Toward-Zhang2023}.
The combined dataset covers diverse input conditions as shown in Table~\ref{tab:corpora}, and more information can be found in~\cite{Toward-Zhang2023}.
Performances are individually reported for each dataset's evaluation partition.

\vspace{-12pt}
\subsection{Configurations}
\label{ssec:conf}
\vspace{-5pt}
In all experiments, the \mathsymbol{USES2-Swin} / \mathsymbol{USES2-Comp} model comprises $\mathsymbol{K}=3$ / $\mathsymbol{K}=4$ multi-path blocks in Fig.~\ref{fig:overview} (b) / (c), where the first $\mathsymbol{K_s}=2$ blocks include the channel modeling module {\hypersetup{hidelinks}\hyperlink{figure:1}{\Circled{\footnotesize 6}}} in Fig.~\ref{fig:overview} (a).
There are totally 12 transformer layers in both models.
STFT\,/\,iSTFT window and hop sizes are respectively set to 32 and 16 ms, regardless of the sampling frequency.
Existing sub-modules' hyperparameters, including embedding dimension, bottleneck dimension, and transformer layer architecture, are identical to \cite{Toward-Zhang2023}.
By default, we take the first channel as the reference channel when processing multi-channel data.
All experiments were conducted using the ESPnet-SE~\cite{ESPnet_SE-Li2021} toolkit, where the Adam optimizer is used for training. The learning rate increases linearly to \texttt{4e-4} in the first $4000$ steps and then decreases by half when the validation performance does not improve for two consecutive epochs.
Four-second chunks are used during training for efficiency.
We use a batch size of four.
For multi-channel data training, we shuffle the channels of each sample and randomly select up to four channels.
The scale-invariant multi-resolution $L_1$ loss in the frequency domain with a time-domain $L_1$ loss term is adopted~\cite{Towards-Lu2022}.
We set the STFT window sizes of the multi-resolution $L_1$ loss to \{256,\,512,\,768,\,1024\} and the time-domain loss weight to 0.5.

\newpara{Metrics.} We report performances using five metrics below:
wide-band PESQ (PESQ-WB)~\cite{PESQ-Rix2001}, STOI~\cite{STOI-Taal2011}, signal-to-distortion ratio (SDR)~\cite{Performance-Vincent2006}, DNSMOS (OVRL)\footnote{\url{https://github.com/microsoft/DNS-Challenge/blob/master/DNSMOS/DNSMOS/sig_bak_ovr.onnx}}~\cite{DNSMOS_P835-Reddy2022}, and word error rate (WER).
The OpenAI Whisper model\footnote{\url{https://huggingface.co/openai/whisper-large-v2}}~\cite{Whisper-Radford2023} is used for WER evaluation.

\vspace{-10pt}
\subsection{Results analysis}
\label{ssec:results}
\vspace{-5pt}

\newpara{Architecture exploration with a single dataset.} Table~\ref{tab:exp_chime4} first describes our incremental exploration with the proposed techniques on the CHiME-4 dataset.
The objective here is to find a model that performs strongly in both simulated and real conditions.
Rows 1 and 2 serve as baselines. 
Conventional USES~\cite{Toward-Zhang2023} shows severely degraded performance for the real test condition with WER surging to 78.1\%.
We first explore the proposed {\hypersetup{hidelinks}\hyperlink{symbol:decoupled_design}{decoupled channel processing}}, 2-stage training, and channel modeling module through rows 3 to 8. 
It is worth noting that the cascaded structure of channel-wise attention and \mathsymbol{TAC} (Rows 5--6) cannot generalize well to real conditions despite its strong performance on simulation data, as mentioned in Section~\ref{ssec:tac}.
Applying {\hypersetup{hidelinks}\hyperlink{symbol:decoupled_design}{decoupled processing}} and {\hypersetup{hidelinks}\hyperlink{symbol:TAttC}{TA$_{\text{tt}}$C}} for channel modeling in rows 7 and 8 gives a good balance of performance between simulated and real conditions.
All metrics in simulated condition are comparable, while multi-channel performance in real condition improves significantly (WER 78.1\% $\rightarrow$ 28.8\%). 
Rows 9 to 12 demonstrate USES2 performances with and without the proposed 2-stage training. 
Both architectures outperform the USES model, while the proposed \mathsymbol{USES2-Comp} with 2-stage training demonstrates the best performance in the real condition in both DNSMOS and WER.
Note that both proposed models reduce the amount of parameters and computational costs compared to USES.
Since \mathsymbol{USES2-Comp} shows overall better performance than \mathsymbol{USES2-Swin} in Table~\ref{tab:exp_chime4}, we adopt this architecture for the rest experiments.

\begin{table}[t]
    \setstretch{0.9}
    \caption{Performance evaluation on various real recordings.}
    \label{tab:exp_real}
    \centering
    \resizebox{\columnwidth}{!}{%
    \setlength{\tabcolsep}{2pt}
    \begin{tabular}{l cccc cccc}
        \toprule
        \multirow{2}{*}{\textbf{Test set}} & \multicolumn{4}{c}{\cellcolor[HTML]{EEEEEE}\textbf{DNSMOS OVRL} $\uparrow$} & \multicolumn{4}{c}{\textbf{WER (\%)} $\downarrow$} \\
        & \cellcolor[HTML]{EEEEEE}\textbf{noisy} & \cellcolor[HTML]{EEEEEE}\textbf{USES} & \cellcolor[HTML]{EEEEEE}\textbf{{\hypersetup{hidelinks}\hyperlink{symbol:excl}{excl}}} & \cellcolor[HTML]{EEEEEE}\textbf{U2-C} & \textbf{noisy} & \textbf{USES} & \textbf{{\hypersetup{hidelinks}\hyperlink{symbol:excl}{excl}}} & \textbf{U2-C} \\
        \hline
        CHiME-4 (Real, 5ch) & \cellcolor[HTML]{EEEEEE}1.46 & \cellcolor[HTML]{EEEEEE}1.58 & \cellcolor[HTML]{EEEEEE}2.89 & \cellcolor[HTML]{EEEEEE}\textbf{3.08} & \textbf{6.7} & 85.9 & 15.6 & 10.3 \\
        REVERB (Real, 8ch) & \cellcolor[HTML]{EEEEEE}1.57 & \cellcolor[HTML]{EEEEEE}\textbf{3.11} & \cellcolor[HTML]{EEEEEE}2.10 & \cellcolor[HTML]{EEEEEE}3.07 & 5.8 & 5.1 & \textbf{4.9} & 5.1 \\
        \bottomrule
    \end{tabular}%
    }
\end{table}

\newpara{Evaluation of USES2 in diverse conditions.} Table~\ref{tab:exp_real} addresses extensive experimental results on various real datasets.
We train a single \mathsymbol{USES2-Comp} model (denoted as ``U2-C'') with two-stage training on a combined dataset~\cite{Toward-Zhang2023} as introduced in Section~\ref{ssec:db} to evaluate its universality in diverse conditions.
We also train a corpus-exclusive SE model with the  \mathsymbol{USES2-Comp} architecture on each corpus alone (denoted as ``\linkdest{symbol:excl}{}excl'') to validate the efficacy of the ``U2-C'' model.
All models are trained on downsampled 8 kHz data for the sake of efficiency, and then evaluated with the original sampling frequencies.
It can be seen that \emph{the multi-channel SE performance is significantly improved on CHiME-4 real data}, with {\raise.17ex\hbox{$\scriptstyle\sim$}}90\% relative improvement in both DNSMOS OVRL and WER.
This is attributed to the proposed {\hypersetup{hidelinks}\hyperlink{symbol:TAttC}{TA$_{\text{tt}}$C}} module in Section~\ref{ssec:tac} that overcomes the limitation of the vanilla \mathsymbol{TAC} design, which can well handle microphone channels with unevenly distributed energy as in CHiME-4.
Meanwhile, the SE and ASR performances on other real data are comparable, showing the robustness of our proposed \mathsymbol{USES2-Comp} model in unseen scenarios.

Finally, we validate whether our proposed \mathsymbol{USES2-Comp} model can preserve the strong performance on simulated conditions in addition to the generalization in real conditions.
Fig.~\ref{fig:simu_results} shows the average performance over five different datasets as listed in Table~\ref{tab:corpora}.
It can be seen that both USES and U2-C models show comparable performances that outperform the corpus-exclusive models (``excl'').
Combining with the observation in Table~\ref{tab:exp_real}, we believe that the proposed \mathsymbol{USES2-Comp} model can be a drop-in replacement for the conventional USES model with reduced parameters and computational costs.

\vspace{-6pt}
\section{Conclusion}
\label{sec:conclusion}
\vspace{-5pt}
In this paper, we proposed two novel architectures, {\hypersetup{hidelinks}\hyperlink{symbol:USES2-Comp}{USES2-Comp}} and {\hypersetup{hidelinks}\hyperlink{symbol:USES2-Swin}{USES2-Swin}}, for improving the {\protect\hypersetup{hidelinks}\hyperlink{symbol:input_condition_invariant}{input condition invariant}} SE in terms of performance and generalization.
Both USES2 models significantly improved the multi-channel processing capability in real conditions, while preserving the strong performance in simulation conditions.
We further proposed a two-stage training strategy to improve the training efficiency, which also circumvents the data balancing issue between single- and multi-channel data.
Extensive experiments on various public corpora demonstrated the effectiveness of the proposed methods.
In future work, we plan to develop a universal SE model that covers diverse distortions in addition.

\clearpage
\bibliographystyle{IEEEtran}
\bibliography{refs}

\end{document}